\documentclass[amssymb,showpacs,aps,prb,floatfix,twocolumn,superscriptaddress]
{revtex4-1}
\usepackage[pdftex]{graphicx}
\usepackage{dcolumn}% Align table columns on decimal point
\usepackage{bm}% bold math
\usepackage{amsmath}
\usepackage{array}
\usepackage{color}
\usepackage{float}
\usepackage{subfigure}
\usepackage{dsfont}
\usepackage{txfonts}
\usepackage{wasysym}
\usepackage{multirow}
\usepackage{sidecap}
\usepackage{xcolor,cancel}
\usepackage[normalem]{ulem}
\usepackage{amsfonts}
\usepackage{amssymb}
\usepackage{bbm}
\usepackage[bbgreekl]{mathbbol}
\usepackage[bbgreekl]{mathbbol}

\DeclareSymbolFontAlphabet{\mathbb}{AMSb}
\DeclareSymbolFontAlphabet{\mathbbl}{bbold}
\newcommand{\+}{\dagger}

\newcommand{\dn}{\downarrow}

\newcommand{\e}{\varepsilon}

\newcommand{\wire}{\mathrm{wire}}
\newcommand{\qdot}{\mathrm{dot}}

\newcommand{\dotwire}{\mathrm{dot-wire}}

\newcommand{\up}{\uparrow}

\begin{document}
% \title{Kondo temperature in spin-orbit coupled one dimensional systems}
\title{Kondo effect in a quantum wire with spin-orbit coupling }
\author{G. R. de Sousa}
\affiliation{Instituto de F\'isica, Universidade Federal de Uberl\^andia, 
Uberl\^andia, Minas Gerais 38400-902, Brazil.}
\author{Joelson F. Silva}
\affiliation{Instituto de F\'isica, Universidade Federal de Uberl\^andia, 
Uberl\^andia, Minas Gerais 38400-902, Brazil.}

\author{E. Vernek}
\affiliation{Instituto de F\'isica, Universidade Federal de Uberl\^andia, 
Uberl\^andia, Minas Gerais 38400-902, Brazil.}
\date{\today}
\begin{abstract}
\end{abstract}
% \keywords{}
\date{\today}
\begin{abstract}
The influence of spin-orbit interactions on the Kondo effect has been under 
debate recently. Studies conducted recently on a system composed by an 
Anderson impurity on a 2DEG with  Rashba spin-orbit have been shown that 
it can enhance or suppress the Kondo temperature ($T_{\rm K}$), depending on 
the relative energy level position of the impurity with respect to the 
particle-hole symmetric point. Here we investigate a system composed by a 
single Anderson impurity side-coupled to a quantum wire with  spin-orbit 
coupling (SOC). We derive an effective Hamiltonian in which the Kondo coupling 
is modified by the SOC. In addition, the Hamiltonian contains two 
other scattering terms, the so called Dzaloshinskyi-Moriya interaction, know to 
appear in these systems, and a new one describing processes similar to the 
Elliott-Yafet scattering mechanisms. By performing a renormalization group 
analysis on the effective Hamiltonian, we find that the correction on the 
Kondo coupling due to the SOC favors and enhancement of the Kondo temperature 
even in the particle-hole symmetric point of the Anderson model, agreeing with 
the NRG results. Moreover, away from the particle-hole symmetric point, $T_{\rm 
K}$ always increases with the SOC, accordingly with the previous 
renormalization group analysis.

\end{abstract}

\maketitle

\section{introduction}

The well-known Kondo effect is a many-body dynamical screening of a 
localized magnetic moment by the spins of itinerant electrons that occurs
at temperatures below the so called the Kondo temperature ($T_{\rm 
K}$).\cite{Hewson} Originally observed in bulk magnetic alloys\cite{Kondo} with 
conspicuous transport features, this effect has been extensively studied in few 
magnetic impurities coupled to one\cite{Madhavan,Manoharan,Crommie} and 
two\cite{Giordano,Webb,Iye} dimensional systems. Recently, a number of studies 
has discussed the effect spin-orbit coupling (SOC) on the Kondo effect on two 
dimensional systems. More specifically, the question on how the SOC modifies the 
Kondo effect  in systems with an isolated magnetic impurities has gained more  
attention.\cite{Malecki,Zarea,Zitko,Diego,Wong,Isaev,Avishai, Andergassen} The 
influence of the effect of SOC on the Kondo physics has gained major interest 
because the former has become remarkably attractive  in condensed matter 
systems.\cite{Winkler,Manchon} For example, SOC is the basic ingredient for 
many different phenomena, extending from the spin manipulation in the 
celebrated Datta-Das transistor\cite{Data-Das}  to the more fundamental physics 
as in the quantum spin-hall effect\cite{Bernevig2} and Majorana 
Fermions.\cite{Read}

Since the Kondo effect involves collectively the spins of the itinerant 
electrons, it is not surprising that SOC---that locks the electron spin with 
their momenta---will modify it. In fact, while in Ref.~\onlinecite{Malecki} it 
was found no change in the Kondo temperature with SOC, recent  
studies\cite{Zarea,Zitko,Diego,Wong} have found a change in the Kondo 
temperature Rashba SOC. Apart from the Ref.~\onlinecite{Diego} that addresses 
the  Kondo effect in graphene, the other ones report arguable results about 
similar systems. On the one hand in  Ref.~\onlinecite{Malecki} it was found that 
the Rashba SOC causes essentially no effect on $T_{\rm K}$. On the other, in  
Ref~\onlinecite{Zarea}, by renormalization group analysis (RGA) and in 
Refs.~\onlinecite{Zitko} and \onlinecite{Wong}, using the numerical 
renormalization group (NRG), report $T_{\rm K}$ dependent on the SOC. 
Although, the actual functional dependency obtained by the NRG seems to differ 
from the RGA approach. This controversy can be attributed to the different 
regimes in which the analysis were carried out and  to some approximations made 
in the RGA. We should stress that the Malecki's idea of studying  the effect 
of SOC on $T_{\rm K}$ using a standard Kondo model was incomplete. This 
became apparent in Ref.~\onlinecite{Zarea}, in which it was shown that the 
standard Kondo model does not include all the scattering phenomena in the 
system.

\begin{figure}[b!]
\centering
\subfigure{\includegraphics[clip,width=3.in]{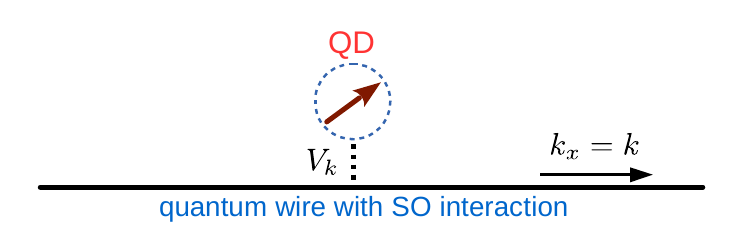}}
\caption{(Color online) Schematic representation of a quantum dot 
side-coupled to a quantum wire with spin-orbit interaction. The wire is 
assumed to lie along the $x$-direction. $V_k$ represents the hopping of 
electrons from the quantum  into the wire.} 
\label{model}
\end{figure} 

Thanks to the various studies discussed above, the effect of SOC on the Kondo 
temperature in two-dimensional systems  have been  quite well elucidated. In 
one dimensional systems, however, the effect of the SOC in the Kondo effect may 
be even more important and has not been investigated so far. The expected 
importance  of the SOC on the Kondo effect on 1D systems can be viewed in as 
simple way. As mentioned above, the Kondo effect is based on scatterings 
accompanied by spin-flip processes  involving the spins of the conduction 
electrons and that one of the local magnetic moments. At very low temperature, 
energy conserving  scatterings become more relevant as compared to 
non-conserving ones. Contrasting with the 2D case, in which energy conserving 
skew scatterings are also allowed, in 1D only forward or backward  scattering 
can occur.  In situations in which  a backward scattering events suffered by 
the conduction electrons requires a flip of their spins, it is expectable that 
the SOC have a much stronger influence in the Kondo effect in 1D systems as 
compared to the 2D ones. Such a spin-momentum locking is known to 
occur in strongly spin-orbit coupled 1D system, such as  InSb 
nanowires\cite{Wimmer}  and in 1D edge state of topological 
insulators.\cite{Hasan}

Motivated by the aforementioned peculiarities of the SOC in one-dimensional 
systems, we investigate the Kondo effect of a magnetic impurity side-coupled to 
a quantum wire with both Rashba\cite{Rashba} and Dresselhaus\cite{Dresselhaus} 
SOC. For the impurity, we restrict ourselves to a spin-1/2 magnetic 
moment and model it as a single level interacting quantum dot that couples to 
the conduction electrons in the quantum wire through tunneling matrix elements. 
By projecting the total Hamiltonian of the system onto a singly occupied 
subspace of the impurity we derive an effective Kondo Hamiltonian, which 
contains the know Dzyaloshinskii-Moriya interaction terms and an additional 
term, analogous to the Elliott-Yafet spin-flip scattering mechanism induced by 
the SOC.\cite{Elliott,Yafet,Fert,Batley} Once we have obtained our effective 
Kondo-like Hamiltonian, we  perform a renormalization group analysis 
(similarly to what was done in Ref.~\onlinecite{Zarea}) from which we extract 
the Kondo temperature. 

Our results show that the dependence of $T_{\rm K}$ with the SOC 
strength  differs from what was found in Ref.~\onlinecite{Zarea}. For instance, 
we find that the Kondo temperature  always increases, even when the system is 
at the  particle-hole symmetric point, that contrats with the results 
reported in Ref.~\onlinecite{Zarea} but agree with those found in 
Refs~\onlinecite{Zitko} and \onlinecite{Wong}. The disagreement between our 
results and those of Ref.~\onlinecite{Zarea} is attributed to the 
correction on the effective Kondo coupling due to the SO interaction,
neglected in the previous study. It is also noteworthy that the dependence of 
$T_{\rm K}$ with the SO coupling is particle-hole asymmetric. We show that an 
extra scattering term in the effective Hamiltonian is the one responsible for 
breaking the particle-hole symmetry of the RG equation.         

The remainder of this paper is organized as follows: In Sec.~\ref{Sec:model} we 
present the model Hamiltonian and derive an effective Kondo-like Hamiltonian 
and in Sec.~\ref{Sec:Renormalization} we perform a renormalization group 
analysis with numerical solution. Finaly, in Sec.~\ref{Sec:Conclusion} we 
summarize our mains results. Some of the details of the calculations are shown 
in the appendices.

\section{Hamiltonian model}
\label{Sec:model}  
For the sake of clarity, we schematically represent our system in 
Fig.~\ref{model}, in which the local magnetic moment is modeled by a 
sigle-level quantum dot occupied by one electron. The quantum wire is 
assumed to lie along the $x$-axis and  includes both  Rashba\cite{Rashba} and 
linear Dresselhaus SOC.\cite{Dresselhaus} Because of the dimensionality of the 
wire, both SOCs are treated in the same footing. More  formally, our system is 
described by an Anderson-like model, $H=H_{\wire}+H_{\qdot}+H_{\dotwire}$, 
where 
\begin{equation}
H_{\qdot}=\sum_{s}  \varepsilon_{d}d_{s}^{\dag} d_{s}+ U n_\up n_\dn,
\end{equation}
describes the isolated quantum dot, in which $d^\dagger_{s}$  ($d_s$) creates 
(annihilates) an electron with energy $\e_d$ and spin $s$ in the dot and $U$ is 
the on-site  Coulomb repulsion in the quantum dot. We also have defined the 
number operator $n_{s}=d^\dag_{s}d_s$.   The quantum wire is described by
\begin{equation}\label{H_RD}
H_{\wire}= \sum_k \left[\varepsilon_k\delta_{ss^\prime}
+k\left(\beta\sigma^{x}_{ss^\prime}-\alpha\sigma^{y}_{s 
s^\prime}\right)\right] c^\dagger_ { k s }c_{k s^\prime},
\end{equation}
where $k$ is  the momentum along $x$-axis, 
$\varepsilon_k={\hbar^{2}k^{2}}/{2m^*}$ the kinetic energy with $m^*$ 
representing the  effective mass of the conduction electrons.  The operator 
$c^\dagger_{ks}$ ($c_{ks}$) creates  (annihilates) an electron with momentum $k$ 
and spin $s$ in the wire. The Rashba and the linear Dresselhaus  spin-orbit 
interaction coupling is parametrized by the interaction strength $\alpha$ and 
$\beta$, respectively, and $\sigma^\nu$ (with $\nu=x,y,z $) represents the Pauli 
matrices. Finally,
\begin{eqnarray}
 H_{\dotwire}=\sum_{ks} \left (V_{k}c_{ks}^{\dag} 
d_{s}+ V_{k}^{*} d_{s}^{\dag}c_{ks} \right ) 
\end{eqnarray}
couples the quantum dot to the wire with overlap matrix element $V_{k}$.

We should keep in mind that we aim to deriving an effective Kondo-like 
Hamiltonian by projecting out the empty and the doubly occupied states of the 
quantum dot. Before doing so, we want to bring the full Hamiltonian into the 
the single impurity Anderson model (SIAM) form. To accomplished this,  
we diagonalize $H_{\wire}$ by performing the following rotation in the spin 
space,
\begin{eqnarray}\label{Basis_Transformation}
\begin{pmatrix}
c_{k+} \\ c_{k-}
\end{pmatrix}={\mathcal U}\begin{pmatrix}c_{k\uparrow} \\ c_{k\downarrow}
\end{pmatrix},
\end{eqnarray}
with 
\begin{eqnarray}
{\mathcal U}=\frac{1}{\sqrt{2}}
\begin{pmatrix}
1 & e^{-i\theta} \\
-ie^{i\theta}& i
\end{pmatrix},
\end{eqnarray}
where $\theta=\tan^{-1}(\beta/\alpha)$. Under this transformation, the 
Hamiltonian $H_{\wire}$ acquires the diagonal form
\begin{eqnarray}
\tilde H_{\wire}=\sum_{k h} \e_{kh}c^ 
\dag_{kh}c_{kh},
\end{eqnarray}
in which $h=+,-$ is the helical quantum number and $\varepsilon_{k h}=\hbar 
k^2/2m^*+h|\gamma| k$ with $\gamma =\alpha -i\beta$. By applying the same 
transformation to the quantum dot operators we see that the form of $H_{\qdot}$ 
and $H_{\dotwire}$ remain unchanged.  Therefore,  in the SO basis, the total 
Hamiltonian acquires the SIAM form
\begin{eqnarray}\label{H_Anderson}
\tilde H=\sum_{h}  \varepsilon_{d}d_{h}^{\dag} d_{h}+ U n_+n_- +\sum_{k h} 
\e_{kh}c^\dag_{kh}c_{kh} \nonumber\\
 +\sum_{kh} \left (V_{k}c_{kh}^{\dag} 
d_{h} +  V^*_{k}d^\dag_{h} c_{kh}  \right), \nonumber\\
\end{eqnarray}
where $\e_{kh}=\e_k+h|\gamma|k$. These are the SO bands shown in 
Fig.~\eqref{diagram}(a). We are now ready to derive the effective 
Kondo-like Hamiltonian.

\subsection{The effective Hamiltonian}
Since we are interested in the Kondo regime of the system in which there is a 
magnetic moment localized in the quantum dot, we project the Hamiltonian 
\eqref{H_Anderson} onto the singly occupied subspace of the quantum dot Hilbert 
space. We follow the same strategy described in Hewson's book\cite{Hewson} 
(for details, see the Appendix \ref{derivation_H_eff}). The resulting effective 
Hamiltonian can be written in the form
\begin{eqnarray}\label{H_Kondo_like}
 H_{\rm eff}=H_0+H_{\rm K}+H_{\rm DM}+H_{\rm EY}.
\end{eqnarray}
Here,
\begin{eqnarray}\label{h0}
 H_0=\sum_{k,h}\e_{kh}c^\+_{kh}c_{kh}
\end{eqnarray}
describes the conduction band on the SO basis, 
\begin{eqnarray}\label{H_K}
H_{\rm K}= \sum_{kk^\prime}J_{kk^\prime}\left[\left(c^\+_{k^\prime 
+}c_{k+}-c^\+_{k^\prime -}c_{k-}\right)S_z 
+c^\+_{k^\prime  +}c_{k-}S_-+c^\+_{k^\prime -}c_{k+}S_+\right]\nonumber\\
\end{eqnarray}
describes  the Kondo coupling, in which
\begin{eqnarray}
J_{kk^\prime}&=&V_kV^*_{k^\prime}\frac{A_k+A_{k^\prime}}{2}, 
\end{eqnarray}
with
\begin{eqnarray}
 A_{k}=\frac{\e_k-\e_d}{(\e_k-\e_d)^2-|\gamma|^2k^2}
+ \frac{\e_d+U-\e_k}{(\e_d+U-\e_k)^2-|\gamma|^2k^2} .
\end{eqnarray}
Observe that $J_{kk^\prime}$ depends on the SO coupling $\gamma$. By inspection 
we see in the absence of the spin-orbit interaction ($\gamma=0$) we recover 
the conventional Kondo coupling, for which $A_{k}=(\e_d+U-\e_k)^{-1} 
+(\e_k-\e_d)^{-1}$. 

The last  two terms of the Hamiltonian \eqref{H_Kondo_like} are given by
\begin{eqnarray}\label{H_DM}
 H_{\rm DM}=\sum_{kk^\prime}\Gamma_{kk^\prime}\left(c^\dag_{k^\prime+}c_{k-}S_- 
- c^\dag_{k^\prime-}c_{k+}S_+\right),
\end{eqnarray}
and
\begin{eqnarray}\label{H_EY}
H_{\rm EY}=H_{\rm EY}^{(1)}+H_{\rm EY}^{(2)}.
\end{eqnarray}
In this last expression,
\begin{eqnarray}\label{H_EY1}
 H^{(1)}_{\rm EY}=\sum_{kk^\prime}\Gamma^{(1)}_{kk^\prime} 
(c^\dag_{k^\prime+}c_{k+} +c^\dag_{k^\prime-}c_{k-})S_z,\nonumber\\
\end{eqnarray}
and 
 \begin{eqnarray}\label{H_EY2}
  H^{(2)}_{\rm EY}=\sum_{kk^\prime}\Gamma^{(2)}_{kk^\prime}
  \frac{n_d}{2}(c^\dag_{k^\prime+}c_{k+}-c^\dag_{
 k^\prime-}c_{k-}).
 \end{eqnarray}
The couplings in the Eqs.~\eqref{H_DM}, \eqref{H_EY1} and \eqref{H_EY2} 
can be written as
\begin{eqnarray}
 \Gamma_{kk^\prime}=V_kV^*_{k^\prime}\frac{B^{(+)}_k-B^{(+)}_{k^\prime}}{2},
\end{eqnarray}
 \begin{eqnarray}
 \Gamma^{(1)}_{kk^\prime}=V_kV^*_{k^\prime}\frac{B^{(+)}_k+B^{(+)}_{k^\prime}}{2
 } ,
 \end{eqnarray}
and 
\begin{eqnarray}
\Gamma^{(2)}_{kk^\prime}=V_kV^*_{k^\prime}\frac{B^{(-)}_k+B^{(-)}_{
k^\prime}}{2}.
\end{eqnarray}
Here we have defined
\begin{eqnarray}
 B^{(\pm)}_{k}=\pm|\gamma|k\left[\frac{1}{(\e_k-\e_d)^2-|\gamma|^2k^2} 
 \mp\frac{1}{(\e_d+U-\e_k)^2-|\gamma|^2k^2}\right].\nonumber\\
\end{eqnarray}
%
% \textcolor{red}{and}
% \begin{eqnarray}
%  \tilde{B}_{k}=-|\gamma|k\left[\frac{1}{(\e_k-\e_d)^2-|\gamma|^2k^2} 
% +\frac{1}{(\e_d+U-\e_k)^2-|\gamma|^2k^2}\right].\nonumber\\
% \end{eqnarray}
% %

% 
The Hamiltonian \eqref{H_DM} corresponds to the known  Dzyaloshinskii-Moriya 
interaction while \eqref{H_EY1} and \eqref{H_EY2} describe the Elliott-Yafet 
like processes,\cite{Elliott,Yafet} responsible for spin-flip scatterings of 
the conduction electrons by  the localized magnetic moments.\cite{Wernick} The 
spin-flip processes involved in the Hamiltonian \eqref{H_EY1} and \eqref{H_EY2} 
are not apparent in the SO basis but is clearly seen when these Hamiltonians 
are written in the real spin representation (see Appendix 
\ref{Spin_Base_Transformation}).

\begin{figure}[t!]
\centering
\subfigure{\includegraphics[clip,width=3.in]{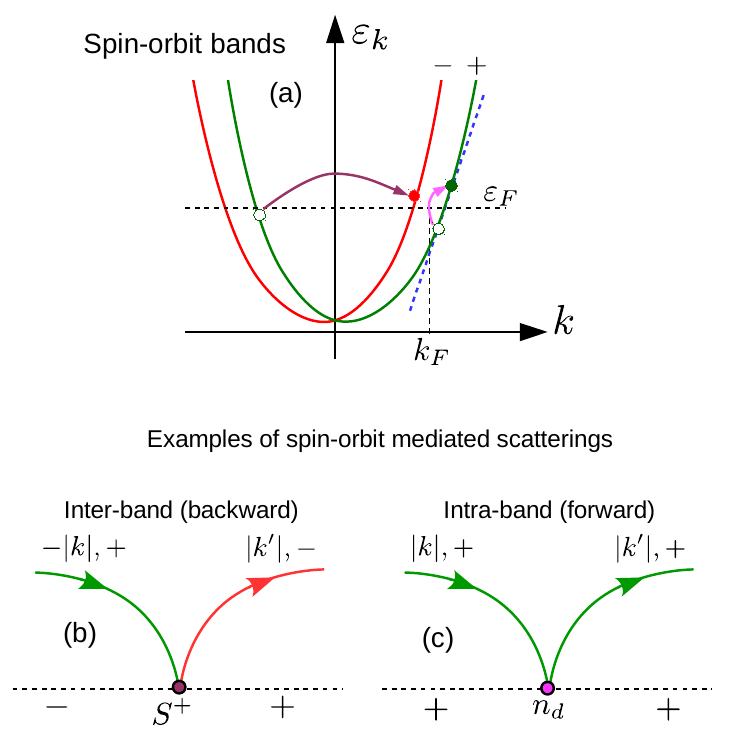}}
\caption{(Color online) (a) Spin-orbit bands for the conduction electrons. At 
low temperature, the allowed processes are those involving energies close to 
the Fermi level $\e_F$. The magenta and purple arrows exemplify, respectively, 
the intra-band (forward) and  intra-band (backward) scatterings. The panels 
(b) and (c) are representative scattering diagrams describing typical 
processes that contained in the Hamiltonians \eqref{H_DM_KF} and 
\eqref{H_EY_KF2}, respectively.} 
\label{diagram}
\end{figure}

At low temperature regime we can assume that the scatterings occurs only 
for electrons with momenta close to Fermi momentum, $k_F$. Moreover, for small 
SO interaction, such that $|\gamma| k_F \ll \hbar k_F^2/2m^*$ (or $|\gamma| \ll 
\hbar k_F/2m^*$), we can set $\varepsilon_{k}\approx \varepsilon_{k_F}=0$ and 
$V_{k}=V_{k_F}\equiv V$. With this we can make the approximations
\begin{eqnarray}\label{J_approx}
J_{kk^\prime}&\approx&|V|^2\left[\frac{\e_d+U}{(\e_d+U)^2-|\gamma_F|^2}
-\frac{\e_d}{\e_d^2-|\gamma_F|^2}\right]\equiv J,
\end{eqnarray}
\begin{eqnarray}\label{Gamma_approx}
\Gamma_{kk^\prime}&\approx&|V|^2|\gamma|\frac{k-k^\prime}{2}\left[\frac{1}{
\e_d^2-|\gamma_F|^2 } -\frac{1}{(\e_d+U-\e_k)^2-|\gamma_F|^2}\right],
\nonumber \\
\end{eqnarray}
\begin{eqnarray}\label{Gamma_1_approx}
 \Gamma^{(1)}_{kk^\prime}&\approx&|V|^2|\gamma|
 \frac{k+k^\prime}{2}\left[\frac{1}{
\e_d^2-|\gamma_F|^2 } -\frac{1}{(\e_d+U-\e_k)^2-|\gamma_F|^2}\right],
\nonumber \\ 
\end{eqnarray}
and 
% % 
 \begin{eqnarray}\label{Gamma_2_approx}
  {\Gamma}^{(2)}_{kk^\prime}&\approx&-|V|^2|\gamma|
  \frac{(k+k^\prime)}{2}\left[\frac{1}{
 \e_d^2-|\gamma_F|^2 } +\frac{1}{(\e_d+U-\e_k)^2-|\gamma_F|^2}\right].
 \nonumber \\
 \end{eqnarray}
In the equations above we have define $\gamma_F=\gamma k_F$. To obtain 
the expressions \eqref{J_approx}-\eqref{Gamma_2_approx} we have replaced $k^2$ 
and $k^{\prime 2}$ by $k_F^2$ but we were careful with the linear terms, keeping 
$k$ and $k^\prime$ intact. This is because  the sums in the Hamiltonian above 
run for positive and negative momenta. Therefore, considering only scatterings 
around $k_F$ we can replace $|k|$ and $|k^\prime|$ by $k_F$ in the couplings 
\eqref{Gamma_approx}-\eqref{Gamma_2_approx}. With this, the factor $k-k^\prime$ 
in the Eq.~\eqref{Gamma_approx} or $k+k^\prime$  in Eqs.~\eqref{Gamma_1_approx} 
 and \eqref{Gamma_2_approx} can be approximated by zero or $\pm 2k_F$, 
depending on the relative sign between  $k$ and $k^\prime$. Bearing this in 
mind, we see that the coupling \eqref{Gamma_approx} contributes only with 
backward scatterings whereas the Eqs.~\eqref{Gamma_1_approx} and 
\eqref{Gamma_2_approx} contribute only with forward scatterings. Explicitly, at 
$k_F$ we can write

\begin{eqnarray}\label{Gamma_F}
\Gamma = V|^2\gamma_F\left[\frac{1}{\e_d^2-|\gamma_F|^2 } 
-\frac{1}{(\e_d+U)^2-|\gamma_F|^2}\right]=-\Gamma_1,
\end{eqnarray}
and 
\begin{eqnarray}\label{Gamma_2F}
\Gamma_2= -|V|^2\gamma_F\left[\frac{1}{\e_d^2-|\gamma_F|^2 } 
+\frac{1}{(\e_d+U)^2-|\gamma_F|^2}\right]. 
\end{eqnarray}

Inserting these  expressions  into Eqs.~\eqref{H_K}, \eqref{H_DM}, 
\eqref{H_EY1}, and  \eqref{H_EY2} we obtain 
\begin{eqnarray}\label{H_K_KF}
H_{\rm K}= J\sum_{kk^\prime}\left[\left(c^\+_{k^\prime 
+}c_{k+}-c^\+_{k^\prime -}c_{k-}\right)S_z +c^\+_{k^\prime  
+}c_{k-}S_-+c^\+_{k^\prime -}c_{k+}S_+\right],\nonumber\\
\end{eqnarray}
\begin{eqnarray}\label{H_DM_KF}
 H_{\rm DM}=\Gamma\sum_{kk^\prime >0}\left(c^ \+_{-k^ \prime+}c_{k-}S_-
- c^\+_{k^\prime +}c_{-k-}S_- +c^\+_{k^\prime-}c_{-k+}S_+ \right.\nonumber\\ 
 - \left.c^\dag_{-k^\prime-}c_{k+}S_+\right),\quad
\end{eqnarray}
% 
% and
% 
\begin{eqnarray}\label{H_EY_KF1}
H_{\rm EY}^{(1)}&=&\Gamma_1 \sum_{kk^\prime >0}\left[S_z(c_{k^\prime 
+}^{\dag}c_{k+}-c_{-k^\prime +}^{\dag}c_{-k+})  \right.\nonumber\\
&&\qquad\qquad\qquad\left.+S_z(c_{k^\prime -}^{\dag}c_{k-}-c_{-k^\prime 
-}^{\dag}c_{-k-})\right]
\end{eqnarray}
 \begin{eqnarray}\label{H_EY_KF2}
 H_{EY}^{(2)}&=&\Gamma_2  \sum_{kk^\prime 
 >0}\left[\frac{n_{d}}{2}(c_{k^\prime +}^{\dag}c_{k+}-c_{k^\prime 
 -}^{\dag}c_{k-}) \right.\nonumber\\
 &&\qquad\qquad \qquad\left.+\frac{n_{d}}{2}(c_{-k^\prime 
-}^{\dag}c_{-k-}-c_{-k^\prime +}^{\dag}c_{-k+})\right].
 \end{eqnarray}
%  
% Now, the couplings $J$, $\Gamma_1$ and $\Gamma_2$ in 
% Eqs.~\eqref{H_K_KF}-\eqref{H_EY_KF2} represent the various couplings evaluated 
% at the Fermi level. 
% %
% \begin{eqnarray}\label{Gamma_F}
% \Gamma = V|^2\gamma_F\left[\frac{1}{\e_d^2-|\gamma_F|^2 } 
% -\frac{1}{(\e_d+U)^2-|\gamma_F|^2}\right]=-\Gamma_1,
% \end{eqnarray}
% %  
% \begin{eqnarray}\label{Gamma_2F}
% \Gamma_2= -|V|^2\gamma_F\left[\frac{1}{\e_d^2-|\gamma_F|^2 } 
% +\frac{1}{(\e_d+U)^2-|\gamma_F|^2}\right]. 
% \end{eqnarray}

Note that it is now explicit  that the processes in the Hamiltonians $H_{\rm 
DM}$ and in $H_{\rm EY}$ involve only backward and forward scatterings, 
respectively. Moreover, we see that the backward scatterings occur are 
inter-band while the forward ones are intra-band scatterings. These backward 
(inter-band)  and forward (intra-band) scatterings are exemplified with the 
diagrams of  Fig.~\ref{diagram}(b) and \ref{diagram}(c). Because of this very 
well defined scattering processes, it is convenient to split the Kondo, 
likewise. Separating the terms of \eqref{H_K_KF} involving definite backward and 
 forward processes as
\begin{eqnarray}\label{Kondo_split}
 H_{\rm K}&=&J^{\rm F}_\parallel\sum_{kk^\prime>0\atop 
kk^\prime<0}\left(c^\+_{k^\prime 
+}c_{k+}-c^\+_{k^\prime -}c_{k-}\right)S_z \nonumber\\
&&+J^{\rm B}_\parallel\sum_{k>0,k^\prime<0\atop 
k<0,k^\prime>0}\left(c^\+_{k^\prime +}c_{k+}-c^\+_{k^\prime -}c_{k-}\right)S_z 
\nonumber\\ 
&&+ J^{\rm F}_\perp\sum_{kk^\prime>0\atop kk^\prime<0}\left[c^\+_{k^\prime 
+}c_{k-}S_-+c^\+_{k^\prime -}c_{k+}S_+ \right]\nonumber\\
&&+ J^{\rm B}_\perp\sum_{k>0,k^\prime<0\atop 
k<0,k^\prime>0}\left[c^\+_{k^\prime 
+}c_{k-}S_-+c^\+_{k^\prime -}c_{k+}S_+ \right].
\end{eqnarray}
As we will see below, because of the SOC, the several Kondo couplings in the
Eq.~\eqref{Kondo_split} will obey different differential equation in the 
renormalization group analysis.

\section{Renormalization group analysis}
\label{Sec:Renormalization}
To study the low-temperature regime of the system we perform a poor-man scaling 
analysis of the effective Hamiltonian \eqref{H_Kondo_like}. We follow the 
original Anderson's approach\cite{Anderson} to obtain the  
renormalization equations for the effective couplings. After a cumbersome but 
straightforward calculation (see Appendix \ref{Scaling}) we find 
\begin{subequations}
\label{RG:equations}
\begin{eqnarray}
\label{J1B} \dot J_{\perp B} &=&-\rho J_{\perp F}J_{\parallel B}-\rho J_{\perp 
B}J_{\parallel F} +\rho \Gamma \Gamma_1 -\rho \Gamma\Gamma_2
\\
\label{J1F} \dot J_{\perp F}&=&-\rho  J_{\perp F}J_{\parallel 
F}-\rho  J_{\perp B}J_{\parallel B}\\
\label{J2B} \dot J_{\parallel B} &=& -2\rho  J_{\perp F}J_{\perp 
B}\\
\label{J2F} \dot J_{\parallel F} &=& -\rho  J_{\perp 
F}^2-\rho J_{\perp B}^2 -\rho  \Gamma^2\\
\label{Gamma} \dot \Gamma  &=& -\rho J_{\parallel F}\Gamma+\rho  J_{\perp B} 
\Gamma_1 -\rho J_{\perp B}\Gamma_2
\\
\label{Gamma_1} \dot \Gamma_1 &=& \rho J_{\perp B} \Gamma + \rho 
J_{\parallel F} \Gamma_2
\\
\label{Gamma_2} \dot \Gamma_2 &=& \rho J_{\parallel F} \Gamma_1.
\end{eqnarray}
\end{subequations}

Following standard notation, in the equations above we have defined $\dot X 
\equiv d X/d\ln \Lambda$, where $\Lambda$ in the reduced bandwidth. We have also 
denoted $\rho =\rho(0)$ as the density of states of the conduction electrons 
calculated and the Fermi level, $\e_F=0$. For this we had to assume that the 
Fermi level is far away above the bottom of the band. In this limit we can 
linearize the band about $k=k_F$ as schematically shown in 
Fig~\eqref{diagram}(a). We can verify that in the absence of SO interaction we 
have the solution for  $\Gamma=\Gamma_1= \Gamma_2=0$, provided the condition has
$\Gamma(D)=\Gamma_1(D) =\Gamma_2(D)=0$. With this, by setting $J_{\perp 
F}=J_{\parallel F}=J_{\perp B}=J_{\parallel B}=J$, the  differential equations 
above reduce to the usual renormalization equation for $J$ in the isotropic 
Kondo model, $\dot J=-2\rho J^2$, leading to the known expression for the Kondo 
temperature, $T_{\rm K}^0=D{\rm Exp}(-1/2\rho  J)$. 

In the presence of SO interaction, an analytical solution for the 
coupled equations \eqref{RG:equations}  is not available. Fortunately,  it can 
be solved  numerically using standard procedures. The numerical solution 
provides us with the coupling as a function of the reduced bandwidth $\Lambda$. 
As in the conventional Kondo model, the Kondo couplings diverge as 
$\Lambda\rightarrow 0$. It is precisely this divergence that provides a 
 definition for the Kondo temperature within the renormalization group analysis. 
Using the same idea here, in  the presence of the SO interaction, we take  as 
$T_{\rm K}$ the value of $\Lambda$ where the numerical solution 
diverges.\footnote{ To check  if this is a good estimation of $T_{\rm K}$ we 
have compared our numerical results with the analytical solution for $\gamma=0$ 
and found a perfect agreement.} 
\begin{figure}[h!]
\centering
\subfigure{\includegraphics[clip,width=3.5in]{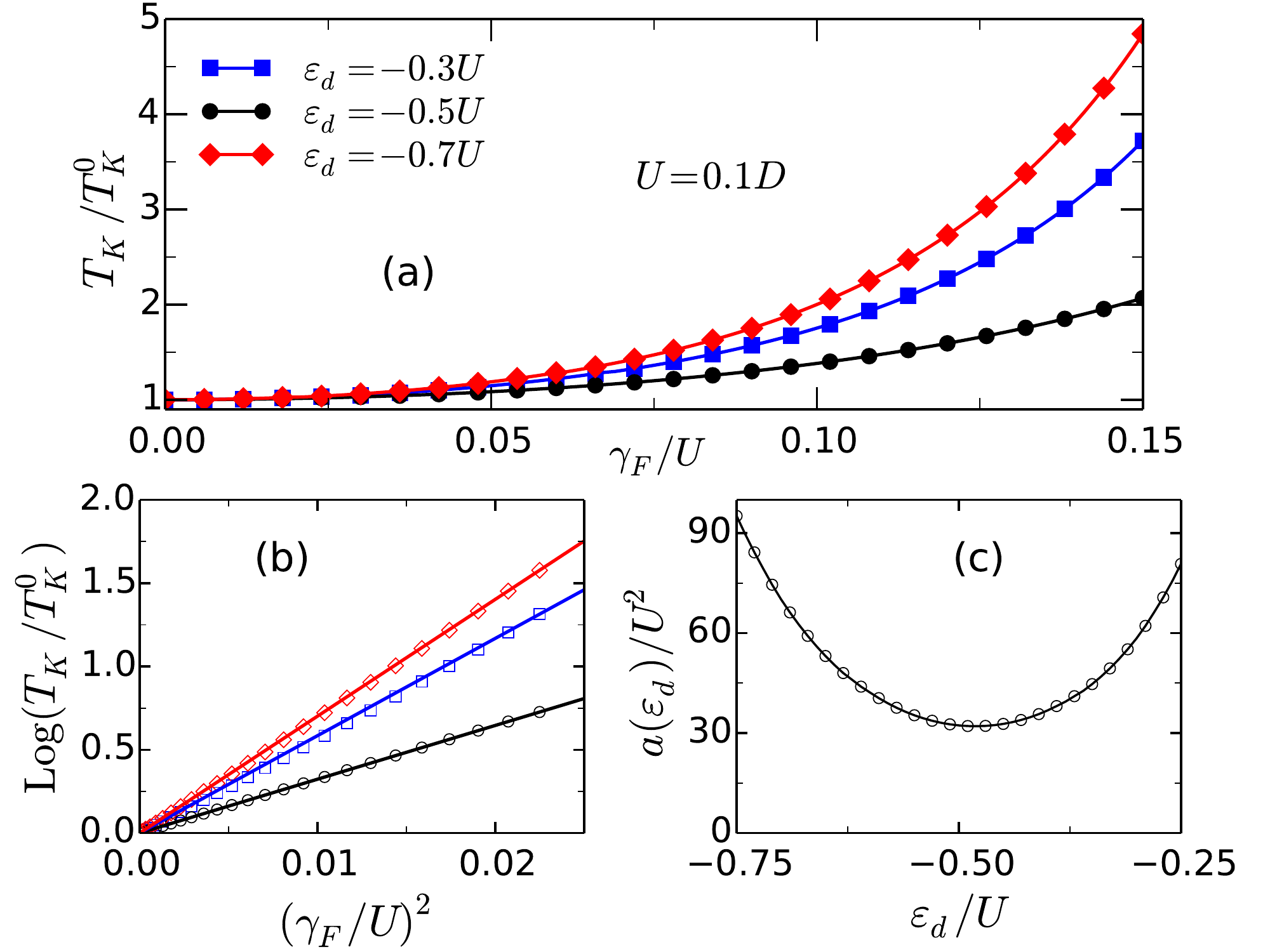}}
\caption{ \label{fig_TK} (Color online) (a) Scaled Kondo temperature vs 
$\gamma_F/U$ for different values of $\varepsilon_d$ and $U=0.1$. 
$\varepsilon_d=-0.5U$ corresponds exactly to the particle-hole symmetric point 
of the Anderson model. Note the different behavior of $T_{\rm K}$ for 
$\varepsilon_d$ above and below $0.05$. $T_{\rm K}^0$ is the Kondo temperature 
calculated in the absence of the SO interaction, $\gamma=0$. (b) ${\rm 
Log}(T_{\rm K}/T_{\rm K}^0)$ vs $\gamma_F/U$ (symbols). The Solid lines show  
straight lines connecting the first and the last points of each data set, 
serving as a guide to the eyes. These lines suggest  that $T_K$ depends on 
$\gamma_F$ exponentially as $T_{\rm K}=T_{\rm K}^0\exp{(a\gamma_F^2)}$, in which 
$a$ is a  function of $\e_d$. (c) $a/U^2$ vs $\e_d/U$ extracted from the results 
of panel (b).}  
\label{kondo_temperature}
\end{figure} 

To obtain our results for $T_{\rm K}$, we set $U/\Delta=20$, with 
$\Delta=\pi V^2/2D$. Here, $D$ is an energy cutoff, within which the band  is 
linearized around $k=k_F$. 
In Fig.~\ref{fig_TK}(a) we show the  Kondo temperature $T_{\rm K}/T_{\rm K}^0$ 
vs $\gamma_F/U$ for three different values of $\e_d$. Here $T_{\rm K}^0$ is the 
Kondo temperature in the absence of the SO interaction. Note that, similarly to 
what was obtained in Ref.~\onlinecite{Zarea}, $T_{\rm K}$ always increases with 
$\gamma_F$, but it is more pronounced  for $\e_d \neq - U/2$ [squares (blue) and 
diamonds (red) curves]. The increase  of $T_{\rm K}$ with $\gamma_F$ for 
$\e_d=-U/2$ contrasts with the results Ref.~\onlinecite{Zarea} that predicts a 
constant $T_{\rm K}$ using the same approach but agrees with those obtained in 
Refs,\onlinecite{Zitko,Liang_Chen,Wong}. The main reason for the disagreement 
with the previous RGA is because they neglected corrections of the Kondo 
coupling $J$ due to the SO interaction. Another compelling point is that for 
$\e_d=-0.7U$ and $\e_d=-0.3U$ for which the impurity level is placed 
symmetrically below and above the particle-hole point, respectively, 
the increasing of $T_{\rm K}$ with $\gamma_F$ is not symmetric. This behavior 
disagree with those of Ref.~\onlinecite{Zarea}. This asymmetry is, however, 
quite different from  asymmetry observed in the results of 
Refs.~\onlinecite{Zitko,Liang_Chen,Wong} because while they considered the 
Fermi level close to the bottom of the conduction band, here we assume $\e_F$ 
far away from it. 

In the absence of analytical solution for the set of differential  
equations \eqref{RG:equations} we attempt to obtain qualitatively the 
dependence of $T_{\rm K}$ on $\gamma_F$. To do so, in 
Fig.~\ref{kondo_temperature}(b) we plot ${\rm Log}(T_{\rm K}/T_{\rm K}^0)$ vs 
$(\gamma_G/U)^2$ for the same three different values of $\e_d$ as in 
Fig.~\ref{kondo_temperature}(a). The symbols correspond the  numerical 
results as shown in \ref{fig_TK}(a) while the solid lines correspond to straight 
lines connecting the first and the last point of the data. Notably, these linear 
functions fit quite well all the data. This suggests a dependence of $T_{\rm K}$ 
on $\gamma_F$ as $T_{\rm K}=T_{\rm K}^0\exp{(a\gamma_F^2)}$,  where $a$ is a 
positive function of the Anderson model parameters (e. g. $\Delta, U, \e_d$). 
Here, by keeping all the other parameters fixed, $a$  clearly shows a strong 
dependence on $\e_d$. To extract a qualitative dependency of  $a$ varies with 
$\e_d$, in Fig.~\ref{kondo_temperature}(c) we  plot $a$ vs $\e_d/U$. Note that 
the shape of the curve is almost parabolic with a minimum close to the 
particle-hole symmetry. It is, However, asymmetric about $\e_d=-U/2$ because of 
the particle-hole asymmetry of the renormalization equation introduced by the 
term $H_{\rm EY}^{(2)}$ of the effective Hamiltonian.

For a better comprehension of the origin of the particle-hole asymmetry in the 
results of Fig.~\ref{fig_TK} let us take a closer look at the renormalization 
equations \eqref{RG:equations}.  We will show that, in fact, the term in the 
Hamiltonian that breaks particle-hole symmetry of the renormalization 
equations is $H_{\rm EY}^{(2)}$, given by the Eq.~\eqref{H_EY_KF2}. To this 
end, let us neglect $H_{\rm EY}^{(2)}$ in the renormalization equations 
Eqs.~\eqref{RG:equations}. We then remove the Eq.~\eqref{Gamma_2} and make 
$\Gamma_2=0$ in all the other equations of the set \eqref{RG:equations}. 
Now, remember that $\Gamma$ and $\Gamma_1$ are odd functions of $\e_d$  
under the change $\e_d=-U/2+\delta$ to $\e_d=-U/2 -\delta$ for any 
$\delta<U/2$.  Therefore, for a given equal initial conditions for  $J$'s 
(which is the case, since $J$ is even) we see that by changing 
 $\e_d=-U/2+\delta$ to $\e_d=-U/2 -\delta$ the derivative of both $\Gamma$ and 
$\Gamma_1$ just change their signs. Now, because the derivatives of the  
$J$'s depends on the product $\Gamma\Gamma_1$ or on $\Gamma^2$, which are 
both even, the resulting value of $T_{\rm K}$ extracted from the solution of 
the Eqs.~\eqref{RG:equations} is particle-hole  symmetric, even though 
$\Gamma$ and $\Gamma_1$ are odd. This show that indeed it is the additional 
term $H^{(2)}_{\rm EY}$ that breaks  the particle-hole symmetry of the 
renormalization equations.
 
\section{Conclusions}
\label{Sec:Conclusion}
Summarizing, we have studied the influence of the Kondo effect of a magnetic 
impurity side coupled to a quantum wire with spin-orbit interaction. We start by
modeling the system with a single impurity Anderson model (SIAM), in which the 
conduction electrons move under both Rashba and Dresselhaus spin-orbit 
couplings. We then derive an effective Kondo model that contains the 
known Dzaloshinskyi-Moriya (DM) interaction and an additional term describing 
scattering processes of the same type of the Elliott-Yafet (EY) mechanisms 
responsible for spin relaxation in systems with magnetic impurities.  We 
splitting the total effective 1D Hamiltonian into forward and backward 
scattering we are able to obtain and then perform a poor-mans scaling to set of 
renormalization equations for the effective couplings. To obtain a Kondo 
temperature dependent of the SO coupling strength we solve numerically 
the coupled equations. We find that the spin-orbit interaction modifies, 
substantially, the Kondo temperature of the system. Our results show that, 
even though the DM term vanishes at the particle-hole (ph) symmetry of the SIAM, 
 and is known to change the Kondo temperature only away from the ph symmetry, 
our study shows that the SOC modifies  the Kondo temperature even in the ph 
symmetry since it modifies the conventional Kondo couplings. Moreover, we find 
that the contribution from additional EY to the enhancement of the  Kondo 
temperature is asymmetric with respect to the ph symmetry. Our study shows 
clearly the scattering mechanisms of the conduction electrons  by the magnetic 
impurity introduced by the SOC in the 1D system. More, importantly, we shown 
how these mechanism change the Kondo temperature of the system. We believe this 
work provides a step forward in the comprehension of the influence of SOC in 
the Kondo effect and is important for future studies, specifically in 1D 
systems.

\acknowledgements
We acknowledge financial support from CNPq, CAPES and FAPEMIG. We would like to 
thank Gerson J. Ferreira for helpful discussions.

\appendix
\section{Derivation of the effective Hamiltonian}
\label{derivation_H_eff}
In order to project the total Hamiltonian \eqref{H_Anderson}  onto the singly 
occupied impurity subspace, we define the projector operators 
\begin{eqnarray}
P_0&=&(1-d^\dagger_\up d_\up)(1-d^\dagger_\dn d_\dn),\\
P_1&=&d^\dag_\up d_\up+d^ \dag_\dn d_\dn-2 d^\dag_\up d^\dag_\dn 
d_\dn d_\up, \\
P_2&=&d^\dag_\up d^\dag_\dn d_\dn d_\up.
\end{eqnarray}
The projected Hamiltonian can be written as
\begin{eqnarray}\label{Projected}
\!H_{\rm 
eff}\!=\!H_{11}\!+\!H_{10}\left(E\!-\!H_{11}\right)^{-1}\!H_{01}\!+\!H_{12}
\left(E\!-\!H_{22}
\right)^{-1}\!H_{21},
\end{eqnarray}
where $H_{ij}=P_iH P_j$. More explicitly, in terms of the creation and 
annihilation  operators, the Hamiltonian \eqref{Projected} can be written as    
\begin{widetext}
\begin{eqnarray}\label{H:full}
H&=&H_0+\sum_{k k^\prime \atop h 
h^\prime}\frac{V_{k^\prime}^*V_k}{2}\Big\{\left[G_h(\e_d,\e_k)+G_{h^\prime}
(\e_d , \e_ { k^\prime } ) 
\right]d_{\bar h}d^\+_{\bar h}d_{\bar h^\prime}d^\+_{\bar h^\prime} d^\+_h 
d_{h^\prime}c_{kh}c^\+_{k^\prime h^\prime} 
-\left[G_h(\e_d+U,\e_k)+G_{h^\prime}
(\e_d +U, \e_ { k^\prime } ) 
\right] n_{d\bar h^\prime} c^\+_{k^\prime 
h^\prime}d_{h^\prime}n_{d\bar h^\prime}d^\+_hc_{kh}\Big\},\nonumber\\
\end{eqnarray}
\end{widetext}
where
\begin{eqnarray}\label{Eq:G}
 G_{h}(\e_d,\e_k)=\frac{1}{\e_d-\e_k}\left[1-\frac{h|\gamma|k}{\e_d-\e_k} 
\right]^{-1}.
\end{eqnarray}
In order to perform the summation on $h$ and $h^\prime$ in the 
Eq.~\eqref{Eq:G} we expand the expression above as a power series of 
$x=h|\gamma|k(\e_d-\e_k)^{-1}$. Summing up the infinite terms of the series we 
can write 

\begin{eqnarray}\label{Eq:G_expanded}
G_{h}(\e_d,\e_k)\!&=&\! \frac{\e_d-\e_k}{(\e_d-\e_k)^2-|\gamma|^2k^2}+\frac{
h|\gamma|k } { (\e_d-\e_k)^2-|\gamma|^2k^2}\nonumber\\
&=& G^{(\rm e)}(\e_d,\e_k) 
+hG^{(\rm o)}(\e_d,\e_k)
\end{eqnarray}
where the first term corresponds to the even order of the series and the 
second on corresponds to the odd terms. We also have used the fact that $h^j=1$ 
for $j$ even and  $h^j=h$ for $j$ odd. It is important to note that the series 
converges only for $|\gamma|k<(\e_d-\e_k)$. This naturally imposes the regime 
of validity of the expansion, $|\gamma|k_F<(\e_d-\e_F)$ and  
$|\gamma|k_F<(\e_d+U-\e_F)$. We can now insert the expression 
\eqref{Eq:G_expanded} into the Eq.~\eqref{H:full} and perform the summation on 
$h$ and $h^\prime$. After lengthy and cumbersome operator algebra we see that 
the even terms will renormalize the Kondo coupling while the odd term will 
provide additional scattering terms in the effective Hamiltonian. The 
resulting Hamiltonian can be split into three terms, namely, $H=H_0+H_{\rm 
K}+H_{\rm DM}+H_{\rm EY}$. The first describes the free conduction electrons 
\begin{eqnarray}
 H_0=\sum_{k,h}\e_{kh}c^\+_{kh}c_{kh},
\end{eqnarray} 
The second term corresponds to the conventional  Kondo Hamiltonian,
\begin{eqnarray}\label{H_K_Appendix}
H_{\rm K}= \sum_{kk^\prime}J_{kk^\prime}\left[\left(c^\+_{k^\prime 
+}c_{k+}-c^\+_{k^\prime -}c_{k-}\right)S_z +c^\+_{k^\prime  
+}c_{k-}S_-+c^\+_{k^\prime -}c_{k+}S_+\right],\nonumber\\
\end{eqnarray}
with a renormalized Kondo coupling,
\begin{eqnarray}
J_{kk^\prime}&=&V_kV^*_{k^\prime}\frac{A_k+A_{k^\prime}}{2}, 
\end{eqnarray}
where 
\begin{eqnarray}
 A_{k}&\!=\!&-G^{(\rm e)}(\e_d,\e_k)+G^{(\rm e)}(\e_d+U,\e_k)\nonumber\\
 &\!=\!&\frac{\e_k-\e_d}{(\e_k-\e_d)^2-|\gamma|^2k^2} + 
\frac{\e_d+U-\e_k}{(\e_d+U-\e_k)^2-|\gamma|^2k^2}. 
\end{eqnarray}

The third  term describes the  Dzaloshinskyi-Moriya scattering 
processes and can be written as
\begin{eqnarray}\label{H_DM_Appendix}
 H_{\rm DM}=\sum_{kk^\prime}\Gamma_{kk^\prime}\left(c^\dag_{k^\prime+}c_{k-}S_- 
- c^\dag_{k^\prime-}c_{k+}S_+\right),
\end{eqnarray}
where the coupling $\Gamma_{kk^\prime}$ is given by
\begin{eqnarray}
 \Gamma_{kk^\prime}&=& V_kV^*_{k^\prime}\frac{B^{(+)}_k-B^{(+)}_{k^\prime}}{2},
\end{eqnarray}
in which we have defined,
\begin{eqnarray}
 B^{\pm}_{k}&=&\pm G^{(\rm o)}(\e_d,\e_k)-G^{(\rm o)}(\e_d+U,\e_k)\nonumber\\
 &=&\pm|\gamma|k\left[\frac{1}{(\e_k-\e_d)^2-|\gamma|^2k^2} 
\mp\frac{1}{(\e_d+U-\e_k)^2-|\gamma|^2k^2}\right].\nonumber\\
\end{eqnarray}

Finally, the fourth term has the form, 
\begin{eqnarray}\label{H_EY_Appendix}
 H_{\rm EY}^{(1)}=\sum_{kk^\prime}\Gamma^{(1)}_{kk^\prime} 
(c^\dag_{k^\prime+}c_{k+}
+c^\dag_{k^\prime-}c_{k-})S_z,\nonumber\\
\end{eqnarray}
 \begin{eqnarray}
 H_{\rm EY}^{(2)}=\sum_{kk^\prime}{\Gamma}^{(2)}_{kk^\prime}
  (c^\dag_{k^\prime+}c_{k+}-c^\dag_{
 k^\prime-}c_{k-})\frac{n_d}{2}
 \end{eqnarray}

with 
\begin{eqnarray}
\Gamma^{(1)}_{kk^\prime}=V_kV^*_{k^\prime}\frac{B^{(+)}_k 
+B^{(+)}_{k^\prime}}{2},
\end{eqnarray}
and
 \begin{eqnarray}
  {\Gamma}^{(2)}_{kk^\prime}= 
 V_kV^*_{k^\prime}\frac{{B}^{(-)}_k+{B}^{(-)}_{k^\prime}}{2},
 \end{eqnarray}

This term has can be thought as describing the Elliott-Yafet-like scattering 
processes in which a electron real spin of the conduction band is flipped upon 
being scattered by the magnetic impurity. This can be better seen if we write 
the Hamiltonian  \eqref{H_EY_Appendix} in the real spin basis, as shown 
in the Appendix~\ref{Spin_Base_Transformation}. 
\section{Real spin representation of the spin-orbit scattering terms}
\label{Spin_Base_Transformation}
It is instructive to see how the effective Hamiltonian looks like in 
the real spin basis. To represent the Hamiltonian back to the real spin basis, 
we use the inverse of the transformation \eqref{Basis_Transformation}. Although 
 this transformation the Kondo Hamiltonian \eqref{H_K} in invariant, the 
spin-orbit scattering terms in the effective Hamiltonian  acquires a different 
form.  After some algebra the spin orbit scattering terms 
\eqref{H_DM_Appendix} and \eqref{H_EY_Appendix} acquire, respectively, the form
\begin{eqnarray}\label{H_DM_Real}
 H_{\rm DM}&=&\frac{i}{2}\sum_{k 
k^\prime}\Gamma_{kk^\prime}\left[\left(c^\+_{k^\prime\up}c_{
k\up} - c^\+_{k^\prime\dn}c_{ 
k\dn}\right)\left(e^{-i\theta}d^\+_{\up}d_{\dn} 
+ e^{i\theta}d^\+_{\dn}d_{\up}\right)\right.\nonumber\\
&&\left.-\left(d^\+_\up d_{\up}-d^\+_\dn d_{\dn} 
\right)\left(e^{-i\theta}c^\+_{k^\prime\up}c_{k\dn}+e^{i\theta}c^\+_{
k^\prime\dn}c_{k\up} \right)\right].
\end{eqnarray}
and 
\begin{eqnarray}\label{H_EY_Real}
 H_{\rm EY}&=&\frac{i}{2}\sum_{k 
k^\prime}\left[\Gamma^{(1)}_{kk^\prime}\left(c^\+_{k^\prime\up}c_{ k\up} + 
c^\+_{k^\prime\dn}c_{ k\dn}\right)\left(e^{-i\theta}d^\+_{\up}d_{\dn} 
-e^{i\theta}d^\+_{\dn}d_{\up}\right)\right.\nonumber\\ 
&& \left.+\Gamma^{(2)}_{kk^\prime}\left(e^{-i\theta}c^\+_{k^\prime\up}c_{ k\dn} 
- e^{i\theta}c^\+_{k^\prime\dn}c_{ k\up}\right)\left(d^\+_{\up}d_{\up} 
+d^\+_{\dn}d_{\dn}\right) \right].\nonumber\\
\end{eqnarray}
The phase factor $e^{\pm \theta}$ appearing in these two last expression can 
be fully gauged  away  by the gauge transformation $c_{k\up}\rightarrow 
e^{-i\theta/2}c_{k\up}$ and $c_{k\dn}\rightarrow e^{i\theta/2}c_{k\dn}$. 
By defining,
\begin{eqnarray}
 {\bf s}_{kk^\prime}=\frac{1}{2}\sum_{ss^\prime} c^\+_{k^\prime 
s}{\bm \tau}_{ss^\prime}c_{k s^\prime} \quad \mbox{and }\quad  {\bf 
S}=\frac{1}{2}\sum_{ss^\prime} d^\+_{s}{\bm \tau}_{ss^\prime}d_{s^\prime},
\end{eqnarray}
with ${\bm \tau}$ being the Pauli matrices including the identity $\tau^0$, we 
can finally  write
\begin{eqnarray}\label{H_DM_Real_Short}
 H_{\rm DM}&=&-2i\sum_{k 
k^\prime}\Gamma_{kk^\prime}\left({\bf s}_{k^\prime k}\times {\bf 
S}\right)\cdot \hat{\bf y},
\end{eqnarray}
which is of the usual form of the Dzaloshinskyi-Moriya interaction, and
\begin{eqnarray}\label{H_EY_Real_Short}
 H_{\rm EY}&=&2\sum_{k k^\prime}\left[\Gamma^{(1)}_{kk^\prime}s^0_{k^\prime 
k}S^y+\Gamma^{(2)}_{kk^\prime}S^0s^y_{k^\prime 
k}\right].
\end{eqnarray}
This expression is similar to the Elliott-Yafet scattering term studied in 
spin relaxation processes.\cite{Fert,Batley}  Note, for instance  that the 
second term contains  spin-flip scattering of the conduction electrons without 
changing the spin of the impurity.

\section{Poor-man scaling analysis}
\label{Scaling}
In the spirit of the  Anderson's perturbative renormalization group, the 
renormalization procedures consists of  progressively reducing the bandwidth of 
 the conduction electrons ($D$) is reduced step-by-step from its initial 
values $D$ towards $D=0$. Within this idea, if at a given step the conduction 
band lies in the interval  $[-\Lambda, \Lambda]$ (where $0<\Lambda\leq D$) it 
is reduced to  $[-(\Lambda+\delta \Lambda), (\Lambda+\delta 
\Lambda)]$ (with $\delta\Lambda <0$) and the part of the Hamiltonian lying 
within the edges of the conduction bands are integrated out while  their 
effects are taken into account perturbatively up to the second order in the 
Hamiltonian coupling.  Using the  $T$-matrix formalism we search for 
scattering processes involving the edge of the conduction bands that 
renormalizes the Hamiltonian, leaving it invariant.~\cite{Anderson} 
Within this idea, if $H_0$ in the unperturbed Hamiltonian and $H_1$ is the 
perturbation, then, up to the second order in the perturbation we can write the 
renormalized interaction by 
\begin{equation}\label{T}
 \tilde H_1=H_1+H_1\frac{1}{E-{H}_{0}}H_1=H_1+\Delta {T}, 
\end{equation}
that has the same form of $H_1$. Note that $\Delta T$ corresponds to the 
change in the $T$-matrix due to all the processes involving the edge 
of the conduction band.   

Explicitly, we can write
\begin{eqnarray}\label{delta_t}
\Delta T & = &\sum_{kk^\prime} \sum_{q \, \mid \, \Lambda-\delta 
\Lambda<\varepsilon_{q}<\Lambda\atop q^\prime \, \mid \, \Lambda-\delta 
\Lambda<\varepsilon_{q^\prime}<\Lambda}   V_{k^{\prime}q^\prime} 
\frac{1}{E-H_{0}}V_{qk} \nonumber \\
& &+\sum_{kk^\prime} \sum_{q \, \mid \, -\Lambda<\varepsilon_{q}<-\Lambda+\delta 
\Lambda \atop q^\prime \, \mid \, -\Lambda<\varepsilon_{q^\prime}<-\Lambda 
+\delta \Lambda} V_{qk} \frac{1}{E-H_{0}} V_{k^{\prime}q^{\prime}},
\end{eqnarray}
Note that in the sum above, $q$ represents momentum such that $\e_q$ lies 
within the edge of the conduction bands. The first term is associated with 
particle states and the second with hole states, removed, respectively, from 
the top and bottom of conduction band. Even though we follow the standard 
procedure found in many textbooks, for the sake of completeness, let us 
illustrate the how term $J_{\parallel B}$ is renormalized  by integrating out 
the degrees of freedom ``living'' at the edge of the conduction band. Using the 
expression~\eqref{delta_t} we see that it rather simple because is not 
renormalized by the SO terms but only by the Kondo coupling terms of the 
Hamiltonian. To shown and example of among the many contribution for the 
Eq.~\eqref{delta_t}, let us calculate product
\begin{eqnarray}\label{exem0}
 H_{\perp F}^{K}\frac{1}{E-H_{0}}H_{\perp B}^{K}
\end{eqnarray}
where $H_{0}$ is given by \eqref{h0} and $H_{\perp \rm F}^{\rm K}$ and $H_{\perp 
\rm B}^{\rm K}$ represent the third and fourth terms of the Hamiltonian 
\eqref{Kondo_split}. Although this term involves only the Kondo coupling, it is 
instructive to show how we deal with the various Kondo couplings split 
into  backward and forward scatterings. 
For the particle-like scattering  processes [first term of the 
Eq.~\eqref{delta_t}] we have 
\begin{eqnarray}
 H_{\perp F}^{K}\frac{1}{E-H_{0}}H_{\perp B}^{K}=J_{\perp F}J_{\perp 
B}\left[\sum_{q^\prime k^\prime>0\atop q^\prime k^\prime<0} 
\left(c^\+_{k^\prime +}c_{q^\prime-} S_-+c^\+_{k^\prime -} 
c_{q^\prime+}S_+\right) \right. \nonumber \\
\left. \times \sum_{k>0,q<0\atop k<0,q>0}\frac{1}{E-H_{0}}\left(c^\+_{q 
+}c_{k-} S_-+c^\+_{q-}c_{k+}S_+\right)\right].\qquad
\end{eqnarray}
Here, we have dropped the constraints for $q$ and $q^\prime$, but recall that 
$q$ and $q^\prime$ run for all momentum such that $\e_{q}$ and $\e_{q^\prime}$ 
lie withing the top edge of the conduction band. Since for a $S=1/2$,  
$S_-^2$ and $S_+^2$ acting on any impurity state vanishes, we can write
\begin{eqnarray}
 H_{\perp F}^{K}\frac{1}{E-H_{0}}H_{\perp B}^{K}=J_{\perp F}J_{\perp B} 
\sum_{q^\prime k^\prime>0\atop q^\prime k^\prime<0}\sum_{k>0,q<0\atop 
k<0,q>0}\left(c^\+_{k^\prime 
+}c_{q^\prime-} \frac{1}{E-H_{0}} \right. 
\nonumber \\
\left.\times c^\+_{q-}c_{k+} S_-S_+ +c^\+_{k^\prime -} 
c_{q^\prime+}\frac{1}{E-H_{0}}c^\+_{q 
+}c_{k-} S_+S_-\right).\qquad
\end{eqnarray}
Using $S_-S_+=1/2-S_z$ e $S_+S_-=1/2+S_z$ and performing the commutations of 
$c^\dagger_{k^\prime +} $ and $c_{q^\prime -}$ with $(E-H_0)^{-1}$ we obtain
\begin{eqnarray}
 H_{\perp F}^{K}\frac{1}{E-H_{0}}H_{\perp B}^{K}=J_{\perp F}J_{\perp B} 
\sum_{q^\prime k^\prime>0\atop q^\prime k^\prime<0}\sum_{k>0,q<0\atop 
k<0,q>0}\left(-\frac{c^\+_{k^\prime 
+}c_{q^\prime-} c^\+_{q-}c_{k+}}{E+\e_{k^\prime+}-\e_{q^\prime-}} 
\right. 
\nonumber \\
\left. +\frac{c^\+_{k^\prime -} c_{q^\prime+}c^\+_{q 
+}c_{k-}}{E+\e_{k^\prime-}-\e_{q^\prime+}}\right)S_z.\qquad
\end{eqnarray}
In the expression above we have neglected the potential scattering term 
generated by the commutations and then set $H_0$ to zero. Now, for the 
top edge (particle-like scattering) we assume $c_{ks}c_{k^\prime 
s^{\prime}}^{\dag}=\delta_{ss^{\prime}}\delta_{kk^\prime}$,  with $s=\pm$. 
Therefore, 
\begin{eqnarray}
 H_{\perp F}^{K}\frac{1}{E-H_{0}}H_{\perp B}^{K}=J_{\perp F}J_{\perp B} 
\sum_{k<0,\, k^\prime,q>0\atop k>0,\, k^\prime,q<0}\left(-\frac{c^\+_{k^\prime 
 +}c_{k+}}{E+\e_{k^\prime+}-\e_{q-}} 
\right. \nonumber \\
\left. +\frac{c^\+_{k^\prime -} c_{k-}}{E+\e_{k^\prime-} 
-\e_{q+}}\right)S_z.\qquad
\end{eqnarray}
Now, since $\e_{qs}$ lies within a very narrow  energy interval near the edge 
of the reduced conduction band we can make $\e_{q+} \sim \e_{q-}\sim \Lambda$ 
to obtain
\begin{eqnarray}
 H_{\perp F}^{K}\frac{1}{E-H_{0}}H_{\perp B}^{K}=-J_{\perp F}J_{\perp B} 
\sum_{k<0,\, k^\prime,q>0\atop k>0,\, k^\prime,q<0}\frac{1} 
{E+\e_{k^\prime+}-\Lambda}\nonumber \\
\times \left(c^\+_{k^\prime +}c_{k+}-c^\+_{k^\prime -} c_{k-} \right) S_z.\qquad
\end{eqnarray}
We now convert the sum in $q$ into integral  and assume a constant density of 
states $\rho$ for the conduction electrons.   Noticing that the sum in $q$ 
is constrained by the sign of $k^\prime$, we can write  
\begin{eqnarray}
 H_{\perp F}^{K}\frac{1}{E-H_{0}}H_{\perp B}^{K}=-J_{\perp F}J_{\perp 
B}\frac{\rho |\delta \Lambda|}{2} \sum_{k<0,\, k^\prime >0\atop k>0,\, 
k^\prime<0}\frac{1}{E+\e_{k^\prime+}-\Lambda}\nonumber \\
\times \left(c^\+_{k^\prime +}c_{k+}-c^\+_{k^\prime -} c_{k-} \right) S_z.\qquad
\end{eqnarray}
For processes near the Fermi level we can neglect $E$ and $\e_{k+}$ in the 
expression, obtaining
\begin{eqnarray}
 H_{\perp F}^{K}\frac{1}{E-H_{0}}H_{\perp B}^{K}\!=\!J_{\perp F}J_{\perp 
B}\frac{\rho |\delta \Lambda|}{2\Lambda} \!\sum_{k<0,\, k^\prime >0\atop k>0,\, 
k^\prime<0} \!\left(c^\+_{k^\prime +}c_{k+}-c^\+_{k^\prime -} c_{k-} \right) 
S_z.\nonumber\\
\end{eqnarray}
Comparing the operators in this expression with those in 
Eq.~\eqref{Kondo_split} we see that this is in fact similar to the second term 
of the Eq.~\eqref{Kondo_split}. Therefore, it contributes to a renormalization 
of $J_{\parallel B}$. Another identical contribution is provided by 
interchanging $H_{\perp F}^{K}$ and $H_{\perp B}^{K}$. Performing the same 
analysis for the  hole-like term in the Eq.~\eqref{delta_t} one finds equal 
contribution. Therefore, the total contribution is given by
\begin{eqnarray}
 \delta J_{\parallel B}=2J_{\perp F}J_{\perp  B}\frac{\rho |\delta 
\Lambda|}{\Lambda} = - 
2J_{\perp F}J_{\perp B}\delta\ln \Lambda. 
\end{eqnarray}
The minor sign in the last step came because $\delta\Lambda <0$. In the limit  
$|\delta \Lambda |\rightarrow 0$ we finally obtain the traditional form 
\begin{eqnarray}
\dot J_{\parallel B}= - 2J_{\perp F}J_{\perp B}. 
\end{eqnarray}

In the calculation above we have considered only two terms of the  the Kondo 
Hamiltonian \eqref{Kondo_split}. Interestingly, after checking all the 
calculation we see that for $J_{\parallel B}$ this is the only contribution. 
Terms involving the SO interaction  will renormalize the other Kondo couplings. 
For example, 

\begin{eqnarray}
 \dot J_{\perp B}=-\rho J_{\parallel F}J_{\perp B} -\rho J_{\parallel 
B}J_{\perp F} +  \rho\Gamma\Gamma_1 -\rho\Gamma\Gamma_2.
\end{eqnarray}

The calculation of all the remaining contributions to the set of 
differential  \eqref{RG:equations}  is lengthy but straightforward.


\begin{thebibliography}{revtex4-1}
% 
\bibitem{Hewson} A.~C.~Hewson,  A. C. Hewson, The Kondo Problem to Heavy 
Fermions (Cambridge University Press, Cambridge, England, 1997).
% 
\bibitem{Kondo} J. Kondo, Prog. Theor. Phys. {\bf 32} (1), 37 (1964).
% 
\bibitem{Madhavan} V. Madhavan, W. Chen, T. Jamneala, M.F. Crommie, and
N. S. Wingreen, Science {\bf 280}, 567 (1998).
% 
\bibitem{Crommie} K. Nagaoka, T. Jamneala, M. Grobis, and M. F. Crommie,
Phys. Rev. Lett. {\bf 88}, 077205 (2002).
% 
\bibitem{Manoharan} G. A. Fiete, J. S. Hersch, E. J. Heller, H.C. Manoharan,
C. P. Lutz, and D.M. Eigler, Phys. Rev. Lett. {\bf 86}, 2392 (2001).
% 
% KOndo in !D 
\bibitem{Giordano} M. A. Blachly and N. Giordano, Phys. Rev. B {\bf 51}, 12537
(1995).
% 
\bibitem{Webb} P. Mohanty and R. A. Webb, Phys. Rev. Lett. {\bf 84}, 4481 
(2000).
%  
\bibitem{Iye} Masahiro Sato, Hisashi Aikawa, Kensuke Kobayashi, Shingo 
Katsumoto, and Yasuhiro Iye, Phys. Rev. Lett. {\bf 95}, 066801 (2005).
% 
% Kondo +  Rashba SOC
\bibitem{Malecki} J. Malecki, J. Stat. Phys. {\bf 129}, 741 (2007).
% 
\bibitem{Zarea} M. Zarea, S. E. Ulloa, and N. Sandler, Phys. Rev. Lett. 
{\bf 108}, 046601 (2012). 
% 
\bibitem{Zitko} R. \v Zitko and J. Bon\v ca, Phys. Rev. B {\bf 84}, 193411 
(2011).
% 
\bibitem{Diego} D. Mastrogiuseppe, A. Wong, K. Ingersent, S. E. Ulloa, and N. 
Sandler, Phys. Rev. B {\bf 90}, 035426 (2014).
% 
\bibitem{Wong} A.~Wong, S.~E.~Ulloa, N.~Sandler, and K.~Ingersent, Phys. Rev. B 
{\bf 93}, 075148 (2016).
% 
\bibitem{Isaev} L.~Isaev, D.~F. Agterberg, and I.~Vekhter, Phys. Rev. B  {\bf 
85} , 081107(R) (2012).
% 
\bibitem{Avishai} K.~Kikoin and Y.~Avishai, Phys. Rev. B {\bf 86}, 155129 
(2012).
% 
\bibitem{Andergassen} S. Grap, V. Meden, and S. Andergassen Phys. Rev. B 
{\bf 86}, 035143 (2012).
% 
\bibitem{Winkler}  R. Winkler,  Spin-orbit Coupling Effects in 
Two-Dimensional Electron and Hole Systems. Springer Tracts in Modern Physics 
(Book 191) (2003).
% 
\bibitem{Manchon} A. Manchon, H.~C.~Koo, J.~Nitta, S.~M. Frolov R.~A.~Duine, 
Nat. Materials {\bf 14}, 871 (2015).
% Persistent Spin Helix
\bibitem{Data-Das} S. Datta and B. Das, Appl. Phys. Lett {\bf 56}, 665 (1990).
% 
\bibitem{Bernevig2} B. Andrei Bernevig, Taylor L. Hughes, Shou-Cheng Zhang, 
Science {\bf 314}, 1757 (2006).
% 
\bibitem{Read} N. Read and Dmitry Green, Phys. Rev. B {\bf 61} 10267 (2000).
% 
\bibitem{Wimmer} I.~van Weperen, B.~Tarasinski, D.~Eeltink, V.~S.~Pribiag, 
S.~R.~ Plissard, E.~P.~A.~M.~Bakkers, L.~P.~Kouwenhoven, and M.~Wimmer, Phys. 
Rev. B {\bf 91} 201413 (2015).
% 
\bibitem{Hasan} M.~Z.~Hasan and C.~L.~Kane, Rev. Mod. Phys. {\bf 82}, 3045 
(2010).
%  
\bibitem{Rashba} Y.~A.~Bychkov and E.~I.~Rashba, J. Phys. C {\bf 17}, 6039 
(1984).
% 
\bibitem{Dresselhaus}  G. Dresselhaus, Phys. Rev. {\bf 100}, 580 (1955).
% Elliot-Yafet
% 
\bibitem{Elliott} R.~J.~Elliott, Phys. Rev. {\bf 96}, 266 (1954).
% 
\bibitem{Yafet}  Y.~Yafet, J. Appl. Phys. {\bf 39}, 853 (1968).
% 
% Relaxation
\bibitem{Wernick} A.~C. Gossard, T.~Y. Kometani, and J. H.~Wernick,  J. of 
App. Phys. {\bf 39}, 849 (1968).
% 
\bibitem{Fert} Albert Fert, Jean-Luc Duvail, and Thierry Valet
Phys. Rev. B {\bf 52}, 6513 (1995).
% 
\bibitem{Batley} J. T. Batley, M. C. Rosamond, M. Ali, E. H. Linfield, G. 
Burnell, and B. J. Hickey Phys. Rev. B {\bf 92}, 220420(R) (2015).
% 
% Poor man scaling
\bibitem{Liang_Chen} Liang Chen, Jinhua Sun, Ho-Kin Tang, Hai-Qing Lin, 	
arXiv:1503.00449 (2015).
\bibitem{Anderson} P. W. Anderson, J. Phys. C: Solid State Phys. {\bf 3} 2436 
(1970).


\end{thebibliography}
\end{document}